\documentclass[12pt,preprint]{aastex}

\begin{document}

\title{Triggering Collapse of the Presolar Dense Cloud Core 
and Injecting Short-Lived Radioisotopes with a Shock Wave. 
VI. Protostar and Protoplanetary Disk Formation}

\author{Alan P.~Boss}
\affil{Department of Terrestrial Magnetism, Carnegie Institution for
Science, 5241 Broad Branch Road, NW, Washington, DC 20015-1305}
\email{aboss@carnegiescience.edu}

\begin{abstract}

 Cosmochemical evaluations of the initial meteoritical abundance of the 
short-lived radioisotope (SLRI) $^{26}$Al have remained fairly constant
since 1976, while estimates for the initial abundance of the
SLRI $^{60}$Fe have varied widely recently. At the high end of this range,
$^{60}$Fe initial abundances have seemed to require $^{60}$Fe
nucleosynthesis in a core collapse supernova, followed by incorporation 
into primitive meteoritical components within $\sim$ 1 Myr. This 
paper continues the detailed exploration of this classical scenario, using 
models of the self-gravitational collapse of molecular cloud cores 
that have been struck by suitable shock fronts, leading to the injection 
of shock front gas into the collapsing cloud through Rayleigh-Taylor
fingers formed at the shock-cloud interface. As before, these models are
calculated using the FLASH three dimensional, adaptive mesh refinement (AMR), 
gravitational hydrodynamical code. While the previous models used FLASH 2.5, 
the new models employ FLASH 4.3, which allows sink particles to be 
introduced to represent the newly formed protostellar object. Sink 
particles permit the models to be pushed forward farther in time to 
the phase where a $\sim 1 M_\odot$ protostar has formed, orbited by a 
rotating protoplanetary disk. These models are thus able to define what 
type of target cloud core is necessary for the supernova triggering 
scenario to produce a plausible scheme for the injection of SLRIs
into the presolar cloud core: a $\sim 3 M_\odot$ cloud core rotating 
at a rate of $\sim 3 \times 10^{-14}$ rad s$^{-1}$ or higher.

\end{abstract}

\keywords{hydrodynamics --- instabilities --- ISM: clouds ---
ISM: supernova remnants --- planets and satellites: formation ---
stars: formation}

\section{Introduction}

 The supernova triggering and injection hypothesis for the origin of
solar system SLRIs was advanced by Cameron \& Truran (1977) on
the basis of the strong evidence (Lee et al. 1976) for once-live 
$^{26}$Al in Ca, Al-rich refractory inclusions (CAIs) from the 
Allende meteorite. The short $^{26}$Al half-life of 0.72 Myr 
led Cameron \& Truran (1977) to hypothesize nucleosynthesis
of $^{26}$Al in a core-collapse supernova (CCSN), followed by injection 
into the presolar cloud and rapid incorporation of the $^{26}$Al into
refractory inclusions. The so-called canonical solar system ratio 
of $^{26}$Al/$^{27}$Al $\sim 5 \times 10^{-5}$ found by Lee et al. (1976)
remains unchanged today (e.g., Dauphas \& Chaussidon 2011). $^{26}$Al 
is also synthesized during the Wolf-Rayet wind phase of massive star
evolution, prior to any CCSN explosion, and mixed into the interstellar
medium by the WR star outflow (e.g., Dwarkadas et al. 2017),
so evidence for live $^{26}$Al cannot be unambiguously linked to a 
supernova (SN) shock front. $^{60}$Fe, on the other
hand, is produced in significant amounts by CCSN, but not by WR stars,
leading to $^{60}$Fe abundances having become the key cosmochemical
evidence for a CCSN origin of the SLRIs. Unfortunately, inferred
initial abundances of $^{60}$Fe have varied widely, starting from a ratio 
of $^{60}$Fe/$^{56}$Fe $\sim 5-10 \times 10^{-7}$ (Tachibana et al. 2006), 
to $\sim 1.15 \times 10^{-8}$ (Tang \& Dauphas 2012),
$\sim 7 \times 10^{-7}$ (Mishra \& Goswami 2014),
$\sim 2 - 8 \times 10^{-7}$ (Mishra \& Chaussidon 2014),
$\sim 8 - 11 \times 10^{-7}$ (Mishra et al. 2016), 
$\sim 0.5 - 3 \times 10^{-7}$ (Telus et al. 2018),
and $\sim 3.8 \times 10^{-8}$ (Trappitsch et al. 2018).
This most recent $^{60}$Fe/$^{56}$Fe ratio is low enough that it
need not require the injection of CCSN-derived $^{60}$Fe into either the 
presolar cloud core or the solar nebula (Trappitsch et al. 2018).
Vescovi et al. (2018) noted the Trappitsh et al. (2018) result,
but decided that the initial $^{60}$Fe/$^{56}$Fe ratio is uncertain,
lying somewhere “inside the wide range of $10^{-8}$ to $10^{-6}$“.
Lugaro et al. (2018) concluded that until the $^{60}$Fe/$^{56}$Fe ratio
is “firmly established”, one cannot decide the stellar source
responsible for the SLRIs, and of $^{26}$Al in particular.

 Vescovi et al. (2018) considered the predicted SLRI abundances
produced by CCSN models of rotating and non-rotating stars with masses
in the range of 13 to 25 $M_\odot$. Contrary to the results of Takigawa
et al. (2008) for 20 to 40 $M_\odot$ CCSN stars, Vescovi et al. (2018) 
found that $^{53}$Mn was overproduced compared to $^{60}$Fe, by as much 
as two orders of magnitude. Vescovi et al. (2018) made certain 
assumptions about the mixing and fallback models used in the calculations, 
and the results can be sensitive to these assumptions (Takigawa
et al. 2008). In fact, other recent CCSN models have found that the 
abundances of SLRIs and stable elements
produced varied by factors of $\sim 100$ as the 
physical assumptions were explored (Mosta et al. 2018). In a major
review article, Lugaro et al. (2018) call for “updated and improved 
stellar models for all sites of production”. Clearly there 
is much to be learned about the SLRI abundances produced by CCSN.

 Other scenarios have been advanced to explain the solar system SLRIs
besides the CCSN trigger scenario, as reviewed by, e.g., Boss (2016).
In particular, Young (2014, 2016) argued for the SLRIs being produced
by massive stars, such as Wolf-Rayet stars, which eject their nucleosynthetic
products into the interstellar medium in star-forming regions. This suggestion
has received support from the recent work of Fujimoto et al. (2018), who
modeled star formation on the scale of the entire Milky Way galaxy. Fujimoto et al. 
(2018) found that their models produced stars with SLRI ratios in the ranges of 
$\sim 10^{-8}$ to $\sim 10^{-3}$ for $^{26}$Al/$^{27}$Al and
$\sim 10^{-9}$ to $\sim 10^{-5}$ for $^{60}$Fe/$^{56}$Fe. These ranges
encompass the solar system SLRI estimates, and extend far beyond it. The
models apply to stars formed from 740 Myr to 750 Myr in the simulation. 
While suggestive, it is unclear how such a simulation of SLRI abundances
relates to a planetary system formed 4.567 Gyr ago, in a galaxy in a
universe with an age of $\sim$ 13.8 Gyr.

 In spite of the checkered history of the cosmochemical motivation
for the supernova triggering and injection scenario, this paper continues
in a long line of theoretical studies of the scenario employing
detailed multidimensional hydrodynamical models, dating back to EDTONS 
code models by Boss (1995) and VH1 code models by Foster \& Boss (1996, 
1977). Beginning with Boss et al. (2008), the FLASH 2.5 AMR code became the 
workhorse code, due to the computational efficiency of an adaptively 
refined mesh for following a shock front across the computational volume. 
Most of the FLASH models were restricted to optically thin regimes 
($< 10^{-13}$ g cm$^{-3}$), where molecular line cooling is effective 
in shock-compressed regions (Boss et al. 2008, 2010; Boss \& Keiser 
2013, 2014, 2015). This restriction prohibits following the calculations 
to when the central protostar forms, orbited by the forming solar nebula.
Boss (2017) approximated nonisothermal effects in the optically thick regime 
with a simple polytropic ($\gamma = 7/5$) pressure law, which mimicks 
the compressional heating in H$_2$ above $\sim 10^{-13}$ g cm$^{-3}$ 
(e.g., Boss 1980). This paper extends this line of research to its logical 
conclusion, by using the computational artifice of sink particles to 
represent newly formed protostars, thereby side-stepping the burden of
attempting to compute the evolution of the contracting protostar, and
allowing the computational effort to remain focused on the shock-induced
collapse of the portions of the pre-collapse cloud that will form a
protoplanetary disk in orbit about the central protostar/sink particle.

\section{Numerical Hydrodynamics Code}

 The models presented in the previous papers in this series were calculated 
with the FLASH 2.5 AMR hydrodynamics code (Fryxell et al. 2000).
Boss et al. (2010) summarized the implementation of FLASH 2.5 in those
models. Here we restrict ourselves to noting what has been changed
by switching to the use of FLASH 4.3. The primary difference is the use 
of sink particles to represent the newly formed, high density protostar.
Sink particles are the Lagrangian extension of the Eulerian sink cell 
artifice first employed by Boss \& Black (1982). The FLASH implementation 
of sink particles was developed by Federrath et al. (2010) and first included
in the FLASH 4.0 release, though without documentation other than a
reference to the Federrath et al. (2010) paper. Full documentation for
sink particles first appeared in FLASH 4.2. Here we rely on the FLASH 4.3
release, which included several sink particle updates over the previous 
releases.

 Sink particles are intended to represent the microphysics of regions in a 
collapsing gas cloud that have a higher density than can be properly
resolved by the numerical grid, possibly leading to spurious fragmentation
into multiple protostellar objects. Truelove et al. (1997, 1998) first 
introduced the {\it Jeans criterion} as a means to prevent spurious fragmentation 
in isothermal collapse calculations. Boss et al. (2000) further 
demonstrated its utility for nonisothermal calculations. The Jeans 
criterion requires that a Cartesian grid cell size $\Delta x$ should remain
smaller that 1/4 of the local Jeans length 
$\lambda_J = (\pi c_s^2 / G \rho)^{1/2}$, where $c_s$ is the local sound
speed, $G$ is the gravitational constant, and $\rho$ is the gas density.
The FLASH AMR code thus seeks to respect the Jeans criterion by inserting
additional grid points in regions in danger of violating the criterion.
However, this is a losing battle in regions undergoing dynamical 
collapse, and once the AMR code reaches the maximum number of additional
levels of resolution that has been specified, the only recourse is to replace
the mass and momentum of the collapsing region with a sink particle. Besides
possibly avoiding artificial fragmentation, this artifice seeks to keep the 
computational time step from becoming vanishingly small as further
grid refinement is halted.

 In FLASH 4.3, sink particle creation and evolution are controlled
primarily by two parameters: the sink density threshold $\rho_{sink}$ and 
the sink accretion radius $r_{sink}$. A third parameter, the sink 
softening radius, is typically set equal to $r_{sink}$.
The sink density threshold sets a lower limit for regions to be considered
as candidates for sink particle creation, while the accretion radius defines 
the spherical region over which a sink particle is allowed to accrete
gas from the AMR grid, as well as the spherical region considered when a 
sink particle is being considered for creation. The latter action is
only undertaken when a high density region is already on the highest
resolution portion of the AMR grid, has a converging flow and a central
gravitational potential minimum, is gravitationally bound and Jeans-unstable, 
and is not within the distance $r_{sink}$ of another sink particle.
When the sink accretion radius is chosen to be $r_{sink} \sim 2.5 \Delta x$,
and the sink density threshold is chosen to be 
$\rho_{sink} = \pi c_s^2 / 4 G r_{sink}^2$, these two choices imply
that the Jeans length is being resolved by about 5 cells on the finest
level of the AMR grid, thereby satisfying the Jeans criterion.

 Verification of the proper functioning of the sink particles formalism
with FLASH 4.3 was achieved by running the SinkMomTest and SinkRotatingCloudCore 
test cases successfully. FLASH 4.3 uses a default value of 32 cells for the 
Jeans refinement criterion, considerably stricter than the 4 cells
suggested by Truelove et al. (1997, 1998) or the 8 cells employed in
the SinkRotatingCloudCore test case. The default value of 32 cells was
used in the models described below.

 Boss et al. (2008) initiated the use of molecular line radiative 
cooling by H$_2$O, CO, and H$_2$ to control the temperature of compressed 
regions at the shock-cloud interface. This approach has been used in the 
subsequent papers in this series, and continues with the present set of 
models. This cooling, however, is only appropriate in optically thin 
regions, and hence is invalid in high density regions. Boss (2017) extended 
the models to later, higher density phases of evolution by employing a 
barotropic equation of state once densities high enough to produce an optically 
thick protostar arose, i.e., for densities above some critical density 
$\rho_{cr}$ (Boss 1980). The same approach is employed here, in addition
to the sink particle artifice, to handle high density regions. The FLASH 
4.3 code was thus modified to include the molecular line cooling only for
densities less than $\rho_{cr}$. In higher density regions, the gas 
pressure $p$ is defined to depend on the gas density as 
$p \propto \rho^\gamma$, where the barotropic exponent $\gamma$ 
equals 7/5 for molecular hydrogen gas. Boss (2017) tested several different
choices for $\rho_{cr}$ in order to find a value that reproduced 
approximately the thermal evolution found in the spherically symmetrical 
collapse models of Vaytet et al. (2013), who studied the collapse of 
solar-mass gas spheres with a multigroup radiative hydrodynamics code.
Boss (2017) found that $\rho_{cr} \sim 5 \times 10^{-13}$ g cm$^{-3}$
was the best choice for reproducing the Vaytet et al. (2013) thermal 
evolutions. The same choice for $\rho_{cr}$ was thus used for the present
models.

\section{Initial Conditions}

 As in the previous models in this series, the initial conditions
consist of a stationary target molecular cloud core that is about to
be struck by a planar shock wave, similar to that seen in the 
Cygnus loop CCSN remnant shown in Figure 1 of Blair et al. 1999.

In the standard model, the target cloud core and the 
surrounding gas are initially isothermal at 10 K, while the shock front 
and post-shock gas are isothermal at 1000 K. The target clouds consist
of spherical cloud cores with radii of of 0.053 pc and Bonnor-Ebert 
radial density profiles. The central density is varied to produce
initial clouds with masses between 2.2 $M_\odot$ and 6.1 $M_\odot$,
embedded in a background rectangular cuboid of gas with a mass of 
1.3 $M_\odot$ and with random noise in the background density 
distribution. The cloud core is assumed to be in solid body rotation 
about the direction of propagation of the shock wave (the $- \hat y$
direction) at angular frequencies $\Omega_i$ ranging between 
$10^{-14}$ rad s$^{-1}$ and $10^{-13}$ rad s$^{-1}$.
Boss \& Keiser (2015) considered the case of clouds rotating about 
an axis perpendicular to the shock front propagation direction.

 In all of the new models, the initial shock wave was defined to have 
a speed of 40 km s$^{-1}$, a width of $3 \times 10^{-4}$ pc, and a 
density of $7.2 \times 10^{-18}$ g cm$^{-3}$, as assumed by Boss (2017). 
This particular choice is based on previous modeling, which examined
a wide range of shock parameters. E.g., Boss et al. (2010) modeled shock 
wave speeds ranging from 1 km s$^{-1}$ to 100 km s$^{-1}$, and found
that collapse and injection were only possible for shock speeds in the 
range of about 5 km s$^{-1}$ to 70 km s$^{-1}$. Boss \& Keiser (2010) 
studied shocks with a speed of 40 km s$^{-1}$, but with varied shock 
thicknesses, as high as 100 times wider than used here. The goal
was to compare results for relatively thin SN shocks and for the
relatively thick planetary nebula winds of AGB stars. Boss \& Keiser 
(2010) found that thick 40 km s$^{-1}$ shocks shredded the 
target clouds, even with lower shock densities. The SN shock front
parameters assumed here are thus in the middle of the sweet spot 
for simultaneous triggering and injection of SLRIs. It is important
to note that these models refer to the radiative phase of SN shock 
wave evolution, after the Sedov blast wave phase (Chevalier 1974). 
In this phase, the SN shock sweeps up gas and dust while radiatively 
cooling.

  The previous modeling with FLASH 2.5 began with Cartesian 
coordinate grids with six top grid blocks, each with eight cells, 
along the $\hat x$ and $\hat z$ directions, and nine or twelve grid blocks,
each with eight cells, in the $\hat y$ direction, along with three or four 
levels of refinement by factors of two on the initial top grid blocks 
(e.g., Boss 2017). With four levels, the resulting grid cells are cubical  
and have a initial size of $10^{15}$ cm. However, use of the
newly improved multigrid Poisson solver in FLASH 4.3 prohibits
the use of more than one block in each coordinate direction
for the top grid. The resulting need for the use of more initial
levels of refinement in FLASH 4.3 compared to FLASH 2.5 has the benefit
of allowing the AMR grid to insert the initial cells only where they are
needed, which improves the computational efficiency of the models.
Accordingly, the FLASH 4.3 models began with a maximum of six 
levels of refinement on the top grid block, with eight top grid
cells in each coordinate direction. Because the computational volume
is a rectangular cuboid with sides of length $4 \times 10^{17}$ cm in 
$\hat x$ and $\hat z$ and $8.2 \times 10^{17}$ cm in $\hat y$, the
resulting cells are not cubical, being about twice as long
in the $\hat y$ direction as in $\hat x$ and $\hat z$.
With six levels of refinement, the smallest cell size in $\hat x$ and 
$\hat z$ is $4 \times 10^{17}/(2^5 \times 8) = 1.6 \times 10^{15}$ cm
and a size twice as large in $\hat y$. When the refinement level
is increased to seven, the smallest cell size is $0.8 \times 10^{15}$ cm
in $\hat x$ and $\hat z$, similar to the FLASH 2.5 models of Boss (2017).

\section{Results}

 Because the coding approach used in FLASH 4.3 differs in certain ways
from that used in FLASH 2.5, all of the specialized coding
regarding the initial conditions, radiative cooling, and gas pressure
laws employed in the previous FLASH 2.5 models needed to be re-coded in
FLASH 4.3. In order to test that these changes had been properly made,
the first step was to repeat the FLASH 2.5 model O from Boss (2017) with
FLASH 4.3. The FLASH 4.3 model did indeed evolve in essentially the 
same way as the FLASH 2.5, though a quantitative comparison is hindered
by the fact that the initial grid structure and random noise in the
background gas density differ in the two models. Nevertheless, the
testing performed confirmed the reproducibility of the results with the 
new code, in the absence of activating the sink particle mechanism that
was employed in the rest of the new models.

\subsection{Sink particle variations}

 The new models all started with a maximum of six levels of refinement,
and were increased to seven levels after about 0.035 Myr, once a
dense core began to form. FLASH 4.3 inserts new levels of refinement
based on gradients in the density distribution and on the need to resolve
the Jeans length, with the latter coming into play later in the
evolution. As previously noted, the recommended value of 
$r_{sink}$ is 2.5 $\Delta x$ and the recommended value of 
$\rho_{sink}$ is $\pi c_s^2 / 4 G r_{sink}^2$. With six levels of refinement, 
$\Delta x = \Delta z = 1.6 \times 10^{15}$ cm and
$\Delta y = 3.2 \times 10^{15}$ cm. With seven levels, 
$\Delta x = \Delta z = 0.8 \times 10^{15}$ cm and
$\Delta y = 1.6 \times 10^{15}$ cm. 
The suggested values for $r_{sink}$ are thus 2.5 times larger, or
$r_{sink} = 4 \times 10^{15}$ cm, $8 \times 10^{15}$ cm, 
$2 \times 10^{15}$ cm and $4 \times 10^{15}$ cm, respectively.
Given these different minimum grid sizes, 
the implications are that $\rho_{sink}$ should range from 
$2.9 \times 10^{-16}$ g cm$^{-3}$, 
$7.4 \times 10^{-17}$ g cm$^{-3}$,
$1.2 \times 10^{-15}$ g cm$^{-3}$, 
to $2.9 \times 10^{-16}$ g cm$^{-3}$, respectively. These values assume 
$c_s = 2 \times 10^{4}$ cm s$^{-1}$, appropriate for 10 K molecular 
hydrogen gas. For higher temperature gas in the optically thick regions,
the sound speed will increase as the square root of the temperature ($T$),
leading to even higher suggested values for $\rho_{sink} \propto T$. 

 Given this wide range of suggested values for $\rho_{sink}$, depending on 
both the grid direction and phase of evolution, there is no obvious choice for
picking a single value for $\rho_{sink}$ or $r_{sink}$. Hence the first 
set of models was run with variations in the $\rho_{sink}$ and $r_{sink}$
parameters, to determine the sensitivity of the resulting evolutions
to their specifications. Table 1 lists the variations in the initial sink 
particle parameters for models starting with the same initial Bonnor-Ebert 
sphere (2.2 $M_\odot$) used in the previous models (e.g., Boss 2017).

 Table 1 clearly illustrates the importance of the choice of 
$r_{sink}$ for the outcome of the models: the results
ranged from a single sink particle to a spray of multiple sink particles,
with disks eventually forming in each case. For the three models with
$\Omega_i = 10^{-14}$ rad s$^{-1}$, model A produced a single sink
particle, model B produced several hundred, and model C produced several
thousand, as $r_{sink}$ was decreased by a factor of ten between each
model in this sequence. Clearly choosing too small a value for 
$r_{sink}$ means that more sinks will appear than are appropriate, because
$r_{sink}$ is so small that newly formed particles cannot be accreted by
previously formed sink particles. In this case, we know from the previous
studies without sink particles (e.g., Boss 2017), that the correct
result for $\Omega_i = 10^{-14}$ rad s$^{-1}$ 
is the formation of a single sink particle at the location of
the density maximum formed by the shock wave. 

 Figure 1 shows the evolution of the location of the single sink
particle formed in model A, which formed after 0.042 Myr of evolution.
Figure 2 depicts the cloud density distribution at 0.039 Myr, just
before the sink particle formed after the density maximum crossed 
the assumed threshold density of $\rho_{sink} = 10^{-15}$ g cm$^{-3}$.
Figure 3 shows model A near the end of the evolution, as the sink
particle and disk are about to disappear off the bottom edge of the
computational grid. A disk has formed in orbit about the central
protostar, represented by the sink particle, as desired, validating
the choices of $\rho_{sink}$ and $r_{sink}$ for this model. However,
it should be noted that a successful outcome in model A was achieved 
only after the number of levels of refinement was increased from the 
initial six levels to seven levels, shortly before the time shown 
in Figure 2. When the refinement was limited to six levels, three
sink particles formed and orbited around each other as they moved
down the grid, as shown in Figure 4, accreting gas and preventing 
a disk from forming. Clearly care must be taken in increasing
the levels of refinement in order to ensure the correct outcome.
For several models, the number of levels of refinement was increased
to eight as a double check on the results with seven levels.

 The remaining five models in Table 1 all had higher initial rotation
rates, $\Omega_i = 10^{-13}$ rad s$^{-1}$. Models D, E, and F form
a sequence with the same decreasing values of $r_{sink}$ as the
previous models A, B, C, and the basic outcome was identical:
with $r_{sink} = 1.5 \times 10^{15}$ cm, a single sink particle was
formed in model D, but with values 10 and 100 times smaller, 
large numbers of sinks were formed in models E and F. With only 
six levels of refinement, disk formation did not occur in model E, 
but an increase to seven levels did lead to disk formation, as was the
case in model A. Models G and H explored the effects of changing
the value for $\rho_{sink}$. In spite of testing values for $\rho_{sink}$
that were either a factor of ten times higher (model G) than in
model D, or ten times lower (model H), the resulting evolution was
quite similar to that of model D: a single sink particle survived,
at the center of the disk formed by the collapsing cloud core,
provided that seven levels of refinement were employed early in
the evolution.

 These results make clear the need to double check any calculation
using sink particles by increasing the maximum levels of refinement, else 
spurious results might ensue. For these models, seven levels appear to
be needed eventually.
Because of the fact that seven levels of refinement were used in the 
sink-forming phases of evolution, and that $r_{sink}$ should 
be based on the smallest grid spacing, the most reasonable choices 
for the sink particle parameters appear to
be $\rho_{sink} = 1 \times 10^{-15}$ g cm$^{-3}$
and $r_{sink} = 1.5 \times 10^{15}$ cm. These choices were used
in nearly all of the remaining models to be discussed.

\subsection{Target cloud core mass variations}

 Figure 3 suggests that the models discussed so far have reached the
goal of following the evolution far enough to form a rotating disk
in orbit around a sink particle representing the newly formed
protostar. However, the sink particles formed in models 
A, D, G, and H all had masses of only $\sim 0.3 M_\odot$ by the time
the particles exited the lower boundary. The un-accreted mass in the disks
and in the rest of the grid was only $\sim 0.1 M_\odot$. Hence, even
if these models were followed farther in time, the total masses of
the final protostars and disks could not be larger than $\sim 0.4 M_\odot$.
That final mass would be fine for forming a typical T Tauri star, but not
for the protosun and solar nebula. As a result, we now present the 
results of a sequence of models where the initial target cloud
core is assigned a higher central density, resulting in higher
initial masses for the Bonnor-Ebert spheres. Table 2 lists the
initial masses and results for these models, where the density was
increased by a factor of 3 for models I and J, a factor of 2 for 
models K and L, and a factor of 1.5 for models M, N, O, P, and Q.
Note that the resulting initial cloud masses are not increased by
factors of 3, 2, and 1.5, respectively, over that of the standard
target cloud (e.g., Boss 2017), because of the finite density
of the background gas and because the radii of the Bonnor-Ebert
spheres had to be decreased slightly from $1.79 \times 10^{17}$ cm
to $1.75 \times 10^{17}$ cm in order to keep the initial shock front
separated from the increased density target clouds.

 Model I collapsed and formed three sink particles early in the
evolution, and hence that model was terminated before a disk formed.
Model J collapsed and formed a single sink particle, but with a mass
of 3.6 $M_\odot$, accompanied by a disk with a mass of 0.8 $M_\odot$.
Clearly, increasing the target cloud mass by a factor of 3 was excessive,
so models with a factor of 2 increase were investigated next.
Model K collapsed to form a single sink particle with a mass of 1.3 $M_\odot$,
close to the desired value of 1 $M_\odot$, but still too high, considering
that 2.4 $M_\odot$ of gas was still available on the grid for accretion.
Model L collapsed and formed an even higher mass sink particle,
1.6 $M_\odot$, along with a 0.5 $M_\odot$ disk. Evidently, a factor of 
2 increase in the target cloud density was also excessive.

 The final five models had an increase by a factor of 1.5 over the
density of the standard Bonnor-Ebert sphere used in the previous models
(e.g., Boss 2017). Model M collapsed to form a 1.1 $M_\odot$ sink
particle, but without a disk. Similarly, model N, with a slightly
higher rate of initial rotation, formed a 1.0 $M_\odot$ sink particle,
again without a disk, as shown in Figure 5. There was a total of
0.65 $M_\odot$ of gas remaining on the grid at this time (0.081 Myr),
but even the highest density gas close to the sink particle showed
no tendency to form a flattened disk configuration. Apparently a higher
initial rotation rate is necessary for disk formation.

 Model O started with $\Omega_i = 3 \times 10^{-14}$ rad s$^{-1}$ and
produced a 1.05 $M_\odot$ mass sink particle orbited by a disk with
a mass of 0.05 $M_\odot$ (Figure 6), exactly the type of outcome that is a 
reasonable approximation for the early protosun and solar nebula.
Figure 7 shows the midplane of the disk, with the sink particle located
in the center of the disk. Model O thus satisfies the basic criteria
for a configuration that could lead to the formation of our solar system
through the triggered collapse scenario.

 Models P and Q similarly satisfy this basic criterion, forming
roughly solar-mass sink particles orbited by disks with masses
of $\sim 0.07 M_\odot$, showing that with higher initial rotation
rates, and higher sink threshold densities, a similar outcome
results, demonstrating the robustness of the models to changes of
this type. Figures 8 and 9 depict the disk configuration near the
end of model P, as the disk and sink particle near the bottom edge
of the computational grid. High temperatures are restricted to
the shock-cloud interface, as the molecular cooling keeps the
disk at 10 K, given its relatively low maximum density of
$\sim 10^{-15}$ g cm$^{-3}$, well within the optically thin regime.
The sink particle has accreted the higher density gas and grown to a
mass of 0.98 $M_\odot$ by this time, while the disk has a mass
of 0.07 $M_\odot$.

\subsection{Injection efficiency}

 The remaining question about the suitability of these models
concerns the efficiency for injection of SLRIs from the shock front
into the disks formed in models O, P, and Q. The basic result is
evident in Figure 10, which shows the color field density for
model P at the same time and on the same spatial scale as the gas
density cross-section shown in Figure 8. As in all the previous
models, the color field is initiated with unit
density inside the shock front, and is injected by Rayleigh-Taylor (R-T)
instabilities at the shock/cloud interface (e.g., Boss \& Keiser 2013).
As shown by Boss \& Foster (1998), the SLRIs ejected by a
CCSN from deep within the massive star have plenty of time to
catch up with the leading edge of the shock front, as it is slowed
down by the impact on the target cloud. Even before this impact,
however, the lagging SLRIs from deep within the exploding star
will catch up through the snowplowing 
of intervening interstellar medium gas and dust encountered
between the CCSN and the target cloud core, which drastically slows
the shock front speed and dilutes the SLRIs.

 R-T instabilities occur when a low density fluid (here, the shock
front) is accelerated into a higher density fluid (here, the target cloud
core). Given the nearly instantaneous nature of the shock-cloud interaction,
one might prefer to refer to the instability as a Richtmeyer-Meshkov 
instability. However, the astrophysical tradition is to refer to this
interaction as a R-T instability (e.g., Wang \& Chevalier 2001), and 
the papers in this series, starting with Foster \& Boss (1997), have
followed this convention. In addition, as noted by Boss \& Keiser (2012):
"The shock corrugates the surface of the compressed cloud core with
a series of indentations indicative of a R-T instability. Meanwhile,
Kelvin-Helmholtz (KH) instabilities caused by velocity shear form in
the relatively unimpeded portions of the shock front at large impact
parameter compared to the center of the target cloud. The shock front gas
and dust is entrained and injected into the compressed region by the
R-T instability, while the K-H rolls result in ablation and loss of
target cloud mass in the downstream flow." Figure 1 of Boss \& Keiser
(2012) explicitly shows the R-T fingers that appear in these models,
while their Figure 2 demonstrates the three dimensional nature of the
fingers. The K-H rolls are also clearly evident in their Figure 1, as
well as in Figure 1 of Boss \& Keiser (2013). Convergence testing to
verify the reality of the R-T fingers has been performed throughout
this series, starting with Foster \& Boss (1996, 1997), continuing
with the focused studies of Vanhala \& Boss (2000, 2002), and by varying
the number of levels of refinement with the FLASH code (Boss et al. 2010;
Boss \& Keiser 2012, 2014, 2015; Boss 2017). The eight levels
used in some of the Boss (2017) models were equivalent to a uniform
grid with a size of $6144^3$ in the regions of highest resolution.

 Figures 11 and 12 present the gas density and color field
distributions in the disk midplane, where the sink particle is
located. The gas disk has a radius of $\sim 500 AU$, considerably larger 
than might be appropriate even for the solar nebula at this early phase 
of evolution. However, the velocity field at this time (Figure 13) shows
a disk rotational velocity at a radius of about $10^{15}$ cm from the
sink particle of $\sim 2 \times 10^5$ cm s$^{-1}$, significantly less
than the Keplerian orbital velocity at that distance from a solar-mass 
protostar of $\sim 4 \times 10^5$ cm s$^{-1}$. The disk is thus expected
to contract considerably during its subsequent evolution.

 The disk midplane shown in Figure 12 allows subtle 
variations in the color field density to be discerned. These
variations owe their existence to the injection of SLRIs by a small
number of Rayleigh-Taylor fingers (Boss \& Keiser 2014). Figure 12
shows that the ambient color field density in the disk is $\sim 0.033$,
with slightly higher densities in the surrounding cloud as high
as $\sim 0.035$. The color field densities are consistent with
the values found in the previous models, e.g., Boss (2017) found
a value of $\sim 0.046$, while Boss \& Keiser (2014) obtained 
$\sim 0.025$, for the standard model. These injection efficiencies
are on low end of what is needed to be consistent with
cosmochemical estimates of what is needed to explain SLRI abundances
for isotopes derived from pre-supernova stars with masses in the 
range of 20 $M_\odot$ and 25 $M_\odot$ (Takigawa et al. 2008).

 Portegies Zwart et al. (2018) studied the interaction of a nearby
(0.15 - 0.40 pc) massive ($\ge 10 M_\odot$) supernova on an already 
formed circumstellar disk, representing the solar nebula. They found
that the shock wave could explain the truncation of the solar nebula
beyond the Kuiper Belt outer edge of about 50 AU, but also found
that the amount of $^{26}$Al accreted from the supernova shock was 
several orders of magnitude too small to match the canonical
$^{26}$Al/$^{27}$Al ratio. The papers in this series have shown that 
injection into the presolar cloud is a considerably more favorable 
scenario for acquiring CCSN SLRIs than late injection into the solar
nebula.

\section{Discussion}

 As noted in the Introduction, the most recent estimate of the initial 
$^{60}$Fe/$^{56}$Fe ratio may not require the injection of CCSN-derived 
$^{60}$Fe into the presolar cloud core (Trappitsch et al. 2018). However, even if 
correct, the Trappitsch et al. (2018) result need not rule out the CCSN triggering
mechanism for the formation of the solar system. Their initial $^{60}$Fe/$^{56}$Fe
ratio is 20 times lower than the ratio $^{60}$Fe/$^{56}$Fe $= 7.5 \times 10^{-7}$ 
assumed in the SLRI analysis of Takigawa et al. (2008). This ratio falls at the
high end of the range considered by Vescovi et al. (2018). Takigawa et al. (2008)
found that four different SLRI abundances ($^{26}$Al/$^{27}$Al, 
$^{41}$Ca/$^{40}$Ca, $^{53}$Mn/$^{55}$Mn, and $^{60}$Fe/$^{56}$Fe) 
could be roughly matched (within factors of 2) by diluting CCSN 
ejecta in presolar cloud material by dilution factors ($f_0$) ranging between
$1.34 \times 10^{-4}$ and $1.3 \times 10^{-3}$ for 20 $M_\odot$ and 25 $M_\odot$ 
CCSN, respectively.

 As used by Takigawa et al. (2008) and the papers in the present series, 
the dilution factor $f_0 = D$ is defined as the ratio of the amount
of mass derived from the CCSN to the mass that was not ejected by the explosion,
i.e., that of the swept-up interstellar medium and the target presolar cloud. 
Large amounts of dilution then mean small values of $D$, as defined. Boss (2017) 
found that a dilution factor $D \sim 1.4 \times 10^{-4}$ was plausible in CCSN triggering and injection models, provided that the CCSN shock had slowed down by 
a factor of 100 through the snowplowing of intervening molecular cloud gas 
before impacting the target cloud. The explicit goal of the papers in this
series has been to find models consistent with the large values of $D$
derived by Takigawa et al. (2008).

 If the Trappitsch et al. (2018) result holds and reduces the desired value of 
$D$ by a factor of about 20, this reduction could be accommodated in part by a 
shock front that snowplowed more than 100 times its mass (per unit area incident 
on the target cloud), i.e., a shock initially moving faster than the 4000 
km s$^{-1}$ implicitly assumed by Boss (2017). Supernova explosions in massive
stars can eject material at much higher speeds, ranging from $\sim$ 
10,000 km s$^{-1}$ for material deep within the envelope, to as high as
100,000 km s$^{-1}$ for the outer regions (Colgate \& White 1966). As a result,
the standard assumption of a 4000 km s$^{-1}$ initial (pre-snowplow) shock speed 
could be too low by a factor of order 10, and would hence require a proportionate
reduction in $D$. In addition, the presolar cloud may have been hit by a 
SLRI-poor portion of the CCSN shock front, as observations of the Cas A
supernova remnant have revealed clumps of the SLRI $^{44}$Ti that vary by factors 
of 4 in density (Grefenstette et al. 2014). Furthermore, shock speeds in the range
of 10 km s$^{-1}$ to 70 km s$^{-1}$ can trigger collapse (Boss et al. 2010)
but result in injection efficiencies that vary by factors of 3 or more
(Boss \& Keiser 2013, 2015). Hence the hydrodynamical models can support 
triggering and injection of considerably smaller amounts of CCSN material than 
that in the most ideal circumstances, low enough to be consistent with the
results of Trappitsch et al. (2018).

 It should also be noted that the nucleosynthetic yields of specific SLRIs 
derived from a CCSN that are to be expected in a supernova remnant are uncertain.
A precise estimate would require detailed knowledge of the stellar mass fractions 
where specific SLRIs are synthesized, and modeling of the mixing and fallback in 
the CCSN model would be needed to get the correct abundances of all of the SLRIs, 
as was done by Takigawa et al. (2008) and Vescovi et al. (2018). Such 
an exercise is worthwhile to pursue in the light of the new $^{60}$Fe/$^{56}$Fe
ratio found by Trappitsch et al. (2018), but such an exercise is far beyond the 
scope of this paper. Recent three dimensional magnetohydrodynamical models of 
CCSN supernova explosions have shown that the predicted abundances of elements and
SLRIs produced by nucleosynthesis can vary by large factors ($\sim 100$), 
depending on the detailed physics of the model (e.g., Mosta et al. 2018).
The more modest goal of the present series of papers is to investigate the
hydrodynamics of the shock wave triggering and injection hypothesis, without
attempting an element by element match to inferred meteoritical SLRI initial
abundances.

\section{Conclusions}

 The FLASH 4.3 AMR code has allowed the models to use sink particles 
to represent newly formed protostars, thereby avoiding the need to 
resolve the thermal and dynamical evolution of the growing protostar.
Previous work with FLASH 2.5 has constrained the shock wave parameters
that are most conducive to triggering and injection, leading to the fairly
narrow range of values (e.g., Boss \& Keiser 2013) needed for success that has
been considered in the present models.
The models have shown that for the standard shock wave values
considered here (e.g., a 40 km s$^{-1}$ shock front), in order to
produce a central protostar with a mass of $\sim 1 M_\odot$, orbited
by a protoplanetary disk with appreciable mass ($\sim 0.05 M_\odot$),
the target cloud should have a mass of $\sim 3 M_\odot$ and be
rotating with an angular velocity of at least $\sim 3 \times 10^{-14}$
rad s$^{-1}$. While these constraints may seem to be overly specific,
considering the wide array of information available for constraining the
formation of the solar system, ranging from detailed knowledge of our 
planetary masses and orbits, to cosmochemical fingerprints of 
long-decayed nucleosynthetic products, we are able to perform a 
crime scene investigation capable of revealing the source of 
the smoking gun.

 The supernova triggering and injection scenario might seem to be an 
exotic mechanism for initiating star formation in general. However, 
detailed laboratory analyses of the primitive meteoritical materials 
involved in the formation of our solar system remain as the ultimate 
arbiters of the need to consider, or abandon, any scenario requiring
a CCSN shortly before the formation of the CAIs. CCSN are the product 
of massive stars found primarily in OB associations, which
are themselves formed from giant molecular cloud complexes. Given
the short lifetimes of massive stars (a few Myr), CCSN typically 
occur while the OB association is still in the vicinity 
(a few pc) of a molecular cloud complex (Liu et al. 2018).
The fact that about 70 supernova remnants in the Milky Way galaxy are 
thought to be associated with molecular clouds (Liu et al. 2018) implies
that the supernova triggering and injection hypothesis of Cameron \&
Truran (1977) may not be as exotic as it might first appear.
This notion is further supported by observations of 20 clouds in the 
Large Magellanic Cloud (LMC), which are being shocked by the N132D supernova 
remnant (Dopita et al. 2018). These observations are consistent with the
LMC clouds being self-gravitating Bonnor-Ebert spheres, with masses of
a few solar masses, that will be forced into collapse by the shock front,
a situation similar to that considered here.

\acknowledgments

 The referee provided numerous suggestions for improvement.
Michael Acierno and Sandy Keiser provided systems support for the flash 
cluster at DTM, where these calculations were primarily computed, requiring 
the continual use of 160 flash cores (20 nodes) for a two year period. 
The Carnegie memex cluster at Stanford University was used for a few 
models as well. The software used in this work was in large part developed 
by the DOE-supported ASC/Alliances Center for Astrophysical Thermonuclear 
Flashes at the University of Chicago.

\clearpage
\begin{deluxetable}{lcccc}
\tablecaption{Initial conditions and results for the models with varied
sink particle creation parameters: initial
target cloud rotation rates ($\Omega_i$, in rad s$^{-1}$), 
sink density threshold (in g cm$^{-3}$),
sink accretion radius (in cm), and the result of the evolution. \label{tbl-1}}
\tablewidth{0pt}
\tablehead{\colhead{Model} 
& \colhead{$\Omega_i$} 
& \colhead{$\rho_{sink}$} 
& \colhead{$r_{sink}$}
& \colhead{result} }
\startdata

A & $10^{-14}$ & $10^{-15}$ & $1.5 \times 10^{15}$ &  one sink, disk \\

B & $10^{-14}$ & $10^{-15}$ & $1.5 \times 10^{14}$ &  multiple sinks, disk \\

C & $10^{-14}$ & $10^{-15}$ & $1.5 \times 10^{13}$ &  multiple sinks, disk \\

D & $10^{-13}$ & $10^{-15}$ & $1.5 \times 10^{15}$ &  one sink, disk \\

E & $10^{-13}$ & $10^{-15}$ & $1.5 \times 10^{14}$ &  multiple sinks, disk \\

F & $10^{-13}$ & $10^{-15}$ & $1.5 \times 10^{13}$ &  multiple sinks, disk \\

G & $10^{-13}$ & $10^{-14}$ & $1.5 \times 10^{15}$ &  one sink, disk \\

H & $10^{-13}$ & $10^{-16}$ & $1.5 \times 10^{15}$ &  one sink, disk \\

\enddata
\end{deluxetable}

\clearpage
\begin{deluxetable}{lcccccc}
\tablecaption{Initial conditions and results for the models with varied
initial target cloud masses: initial
target cloud rotation rates ($\Omega_i$, in rad s$^{-1}$), 
sink density threshold (in g cm$^{-3}$),
initial cloud mass (in $M_\odot$), final sink particle mass (in $M_\odot$),
final disk mass (in $M_\odot$), and the result of the evolution. \label{tbl-1}}
\tablewidth{0pt}
\tablehead{\colhead{Model} 
& \colhead{$\Omega_i$} 
& \colhead{$\rho_{sink}$} 
& \colhead{$M_{cloud}$} 
& \colhead{$M_{sink}$} 
& \colhead{$M_{disk}$}
& \colhead{result} }
\startdata

I & $10^{-14}$          & $10^{-15}$ & 6.1 & 3.1  & - -  & three sinks, no disk \\

J & $10^{-13}$          & $10^{-15}$ & 6.1 & 3.6  & 0.8  & one sink, disk \\

K & $10^{-14}$          & $10^{-15}$ & 3.7 & 1.3  & - -  & one sink, no disk \\

L & $10^{-13}$          & $10^{-15}$ & 3.7 & 1.6  & 0.5  & one sink, disk \\

M & $10^{-14}$          & $10^{-15}$ & 3.0 & 1.1  & - -  & one sink, no disk \\

N & $2 \times 10^{-14}$ & $10^{-15}$ & 3.0 & 1.0  & - -  & one sink, no disk \\

O & $3 \times 10^{-14}$ & $10^{-15}$ & 3.0 & 1.1  & 0.05 & one sink, disk \\

P & $10^{-13}$          & $10^{-15}$ & 3.0 & 0.98 & 0.07 & one sink, disk \\

Q & $10^{-13}$          & $10^{-14}$ & 3.0 & 1.0  & 0.08 & one sink, disk \\

\enddata
\end{deluxetable}

\begin{figure}
\vspace{-3.0in}
\includegraphics[scale=.80,angle=0]{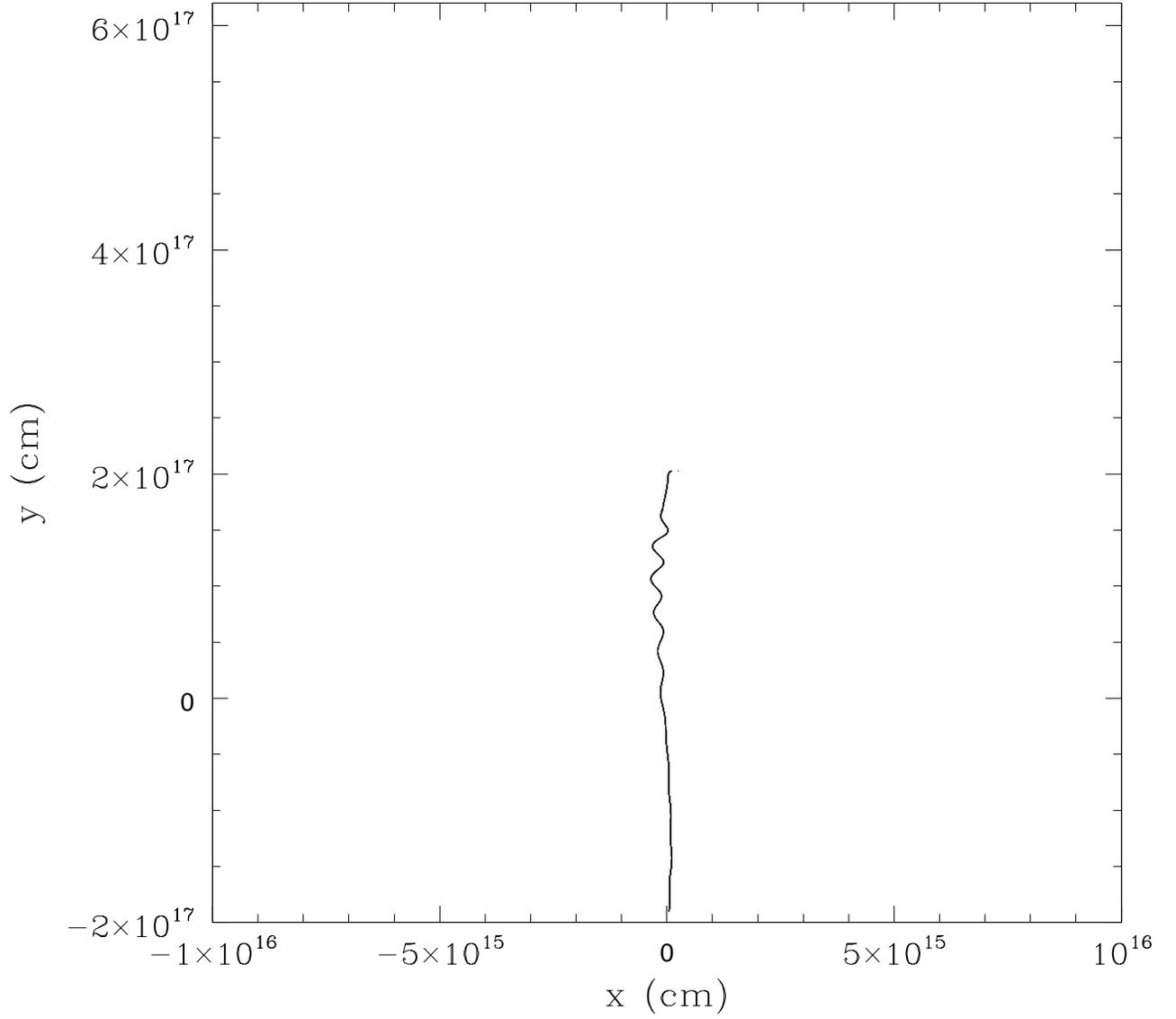}
\vspace{-0.5in}
\caption{Sink particle location in the $z = 0$ plane throughout the evolution 
of model A. The target cloud core is initially centered at $x = 0$, $z = 0$, 
and $y = 4.0 \times 10^{17}$ cm (e.g., see Figure 1 of Boss \& Keiser 2014).
The entire grid along the $y$ axis is shown, but only a small portion (1/20)
of the grid in $x$, for clarity.}
\end{figure}
\clearpage

\begin{figure}
\vspace{-1.0in}
\includegraphics[scale=.80,angle=0]{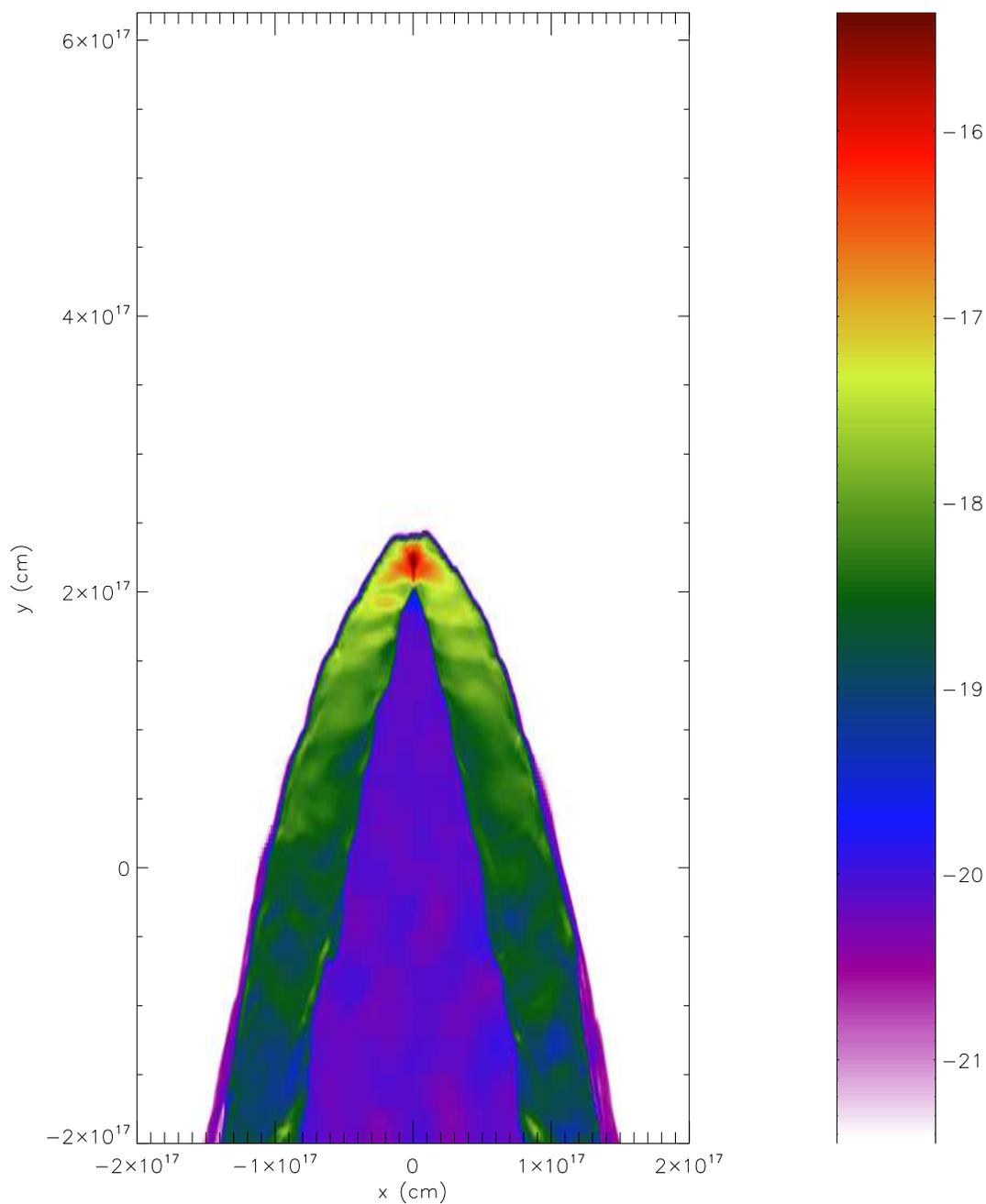}
\vspace{0.5in}
\caption{Model A log density cross-section ($z$ = 0) at 0.039 Myr,
just prior to the formation of the sink particle shown in Figure 1. The 
maximum density is close to the threshold density for sink
particle creation, which is $\rho_{sink} = 10^{-15}$ g cm$^{-3}$
for this model. The entire computational grid is shown, which has
a maximum of seven levels of refinement.}
\end{figure}
\clearpage

\begin{figure}
\vspace{-1.0in}
\includegraphics[scale=.80,angle=90]{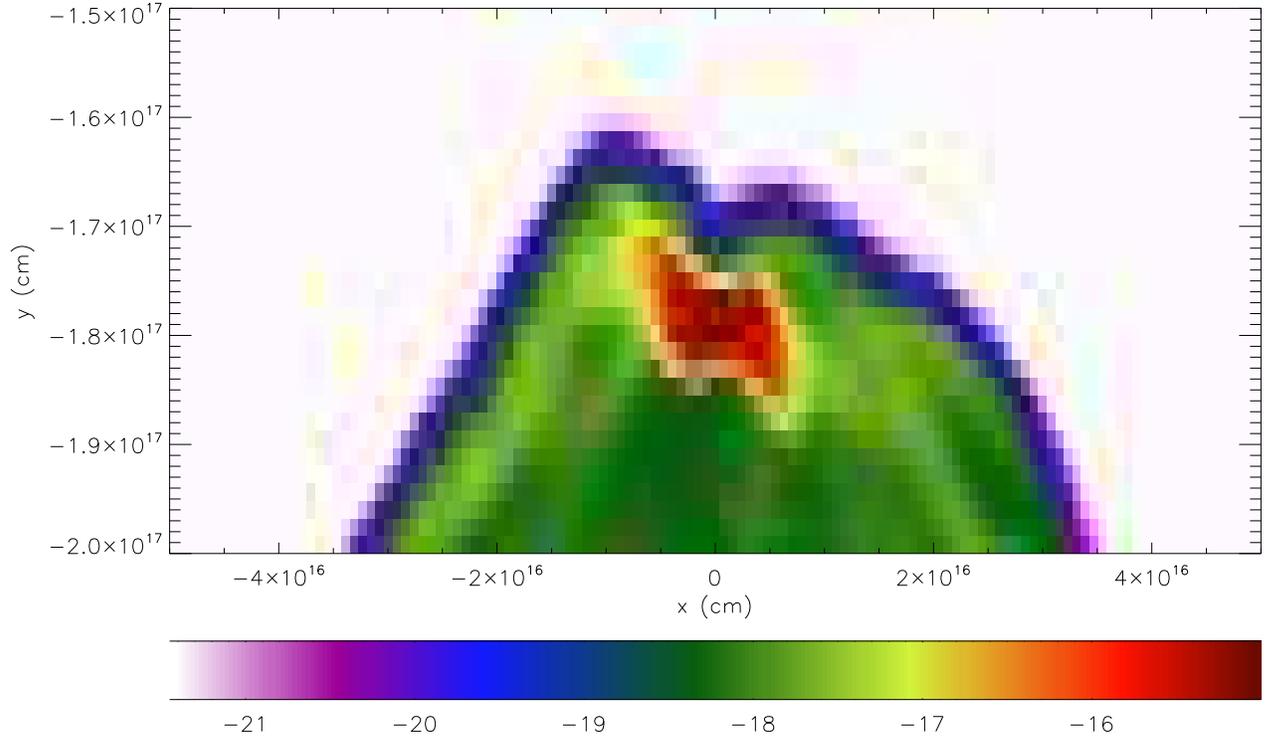}
\vspace{-0.5in}
\caption{Model A log density cross-section ($z$ = 0) at 0.102 Myr. 
At this time, the single sink particle is located at 
$y = -1.79 \times 10^{17}$ cm, in the center of the density maximum
at $x \sim 0$ and $z \sim 0$ (cf., Figure 1). A rotating disk 
has formed around the sink particle, as desired.}
\end{figure}
\clearpage

\begin{figure}
\vspace{-1.0in}
\includegraphics[scale=.80,angle=-90]{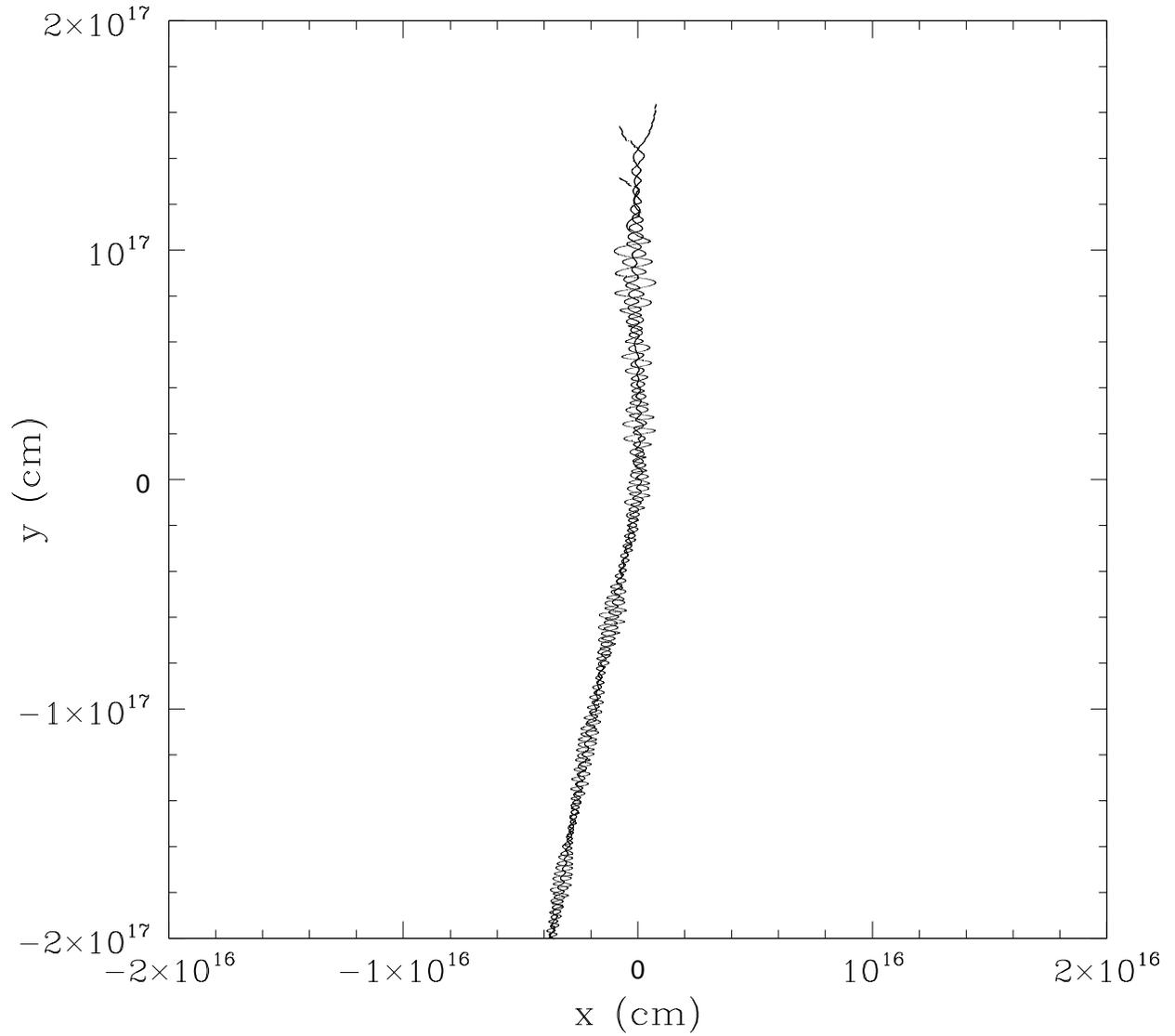}
\vspace{0.5in}
\caption{Sink particle location in the $z = 0$ plane throughout the evolution 
of model A, when only six levels of refinement are allowed. Only
portions of the grids along the $x$ and $y$ axes are shown, for clarity.
The three sink particles accrete gas and prevent the formation of a disk.}
\end{figure}
\clearpage

\begin{figure}
\vspace{-1.0in}
\includegraphics[scale=.80,angle=0]{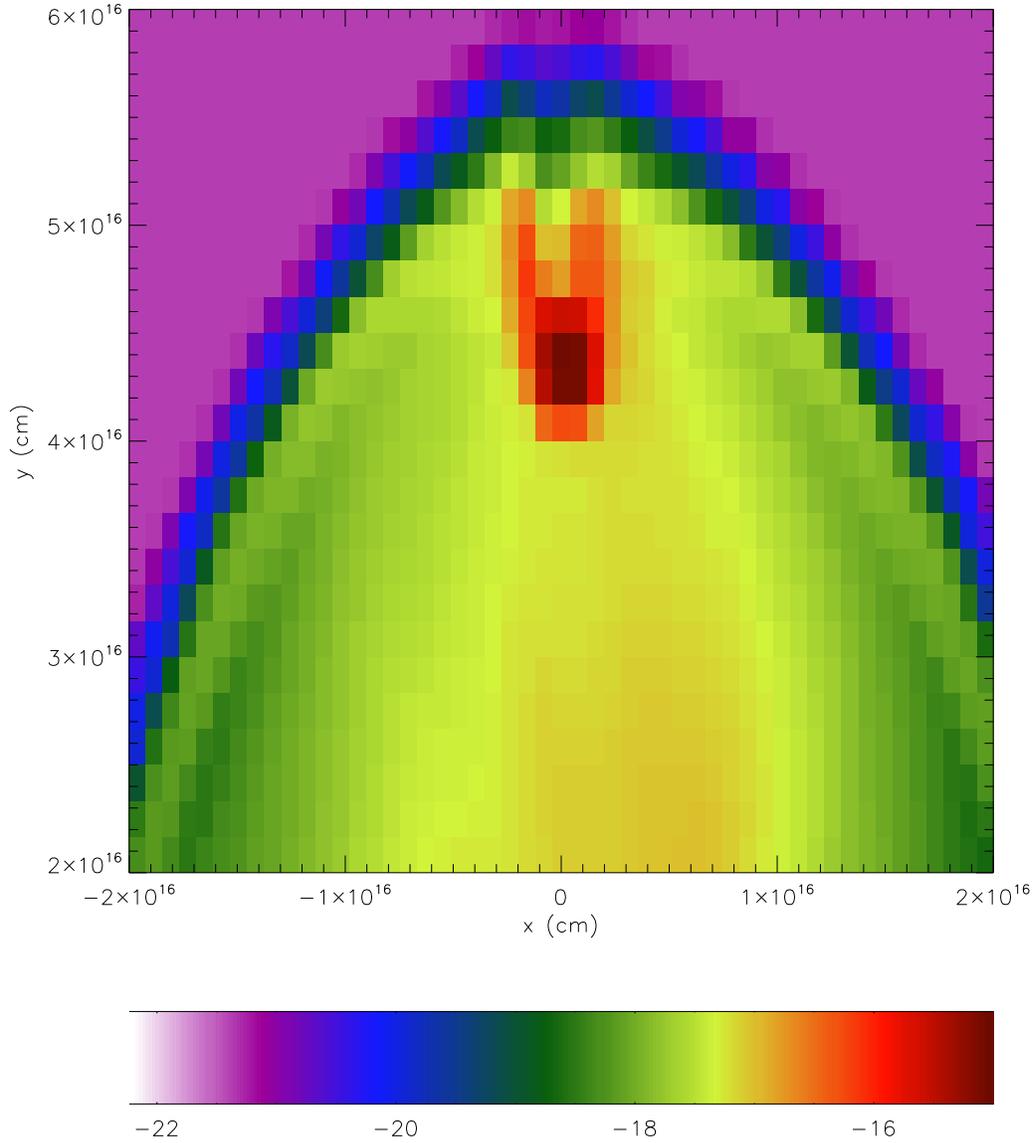}
\vspace{-0.5in}
\caption{Model N log density cross-section ($z$ = 0) at 0.081 Myr. 
At this time, the single sink particle is located at 
$y = 4.2 \times 10^{16}$ cm, inside the densest region. There is
no evidence for disk formation in this model.}
\end{figure}
\clearpage

\begin{figure}
\vspace{-1.0in}
\includegraphics[scale=.80,angle=90]{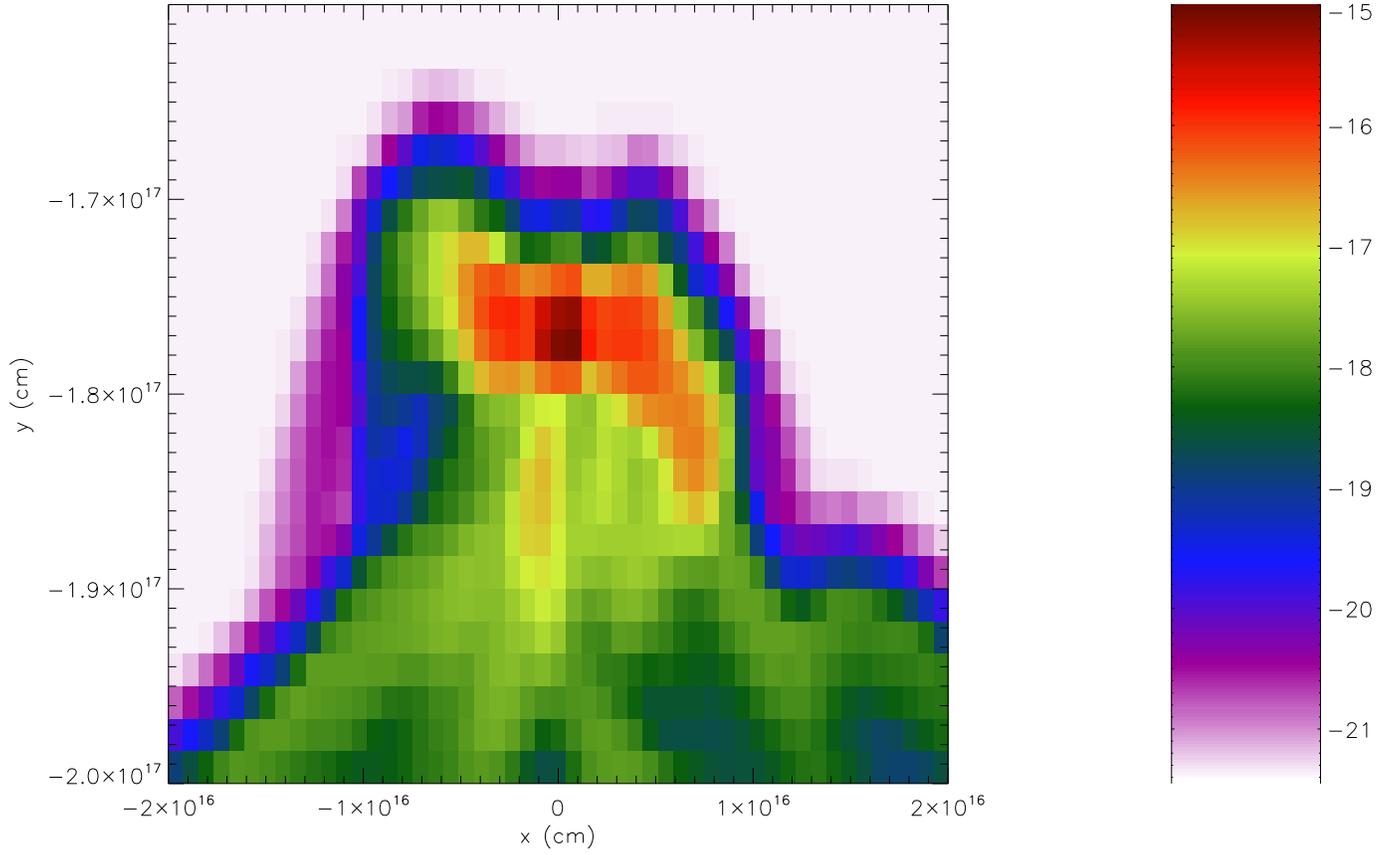}
\vspace{0.5in}
\caption{Model O log density cross-section ($z$ = 0) at 0.121 Myr. 
At this time, the single sink particle is located at 
$y = -1.77 \times 10^{17}$ cm, inside the dense, distorted disk.}
\end{figure}
\clearpage

\begin{figure}
\vspace{-1.0in}
\includegraphics[scale=.70,angle=0]{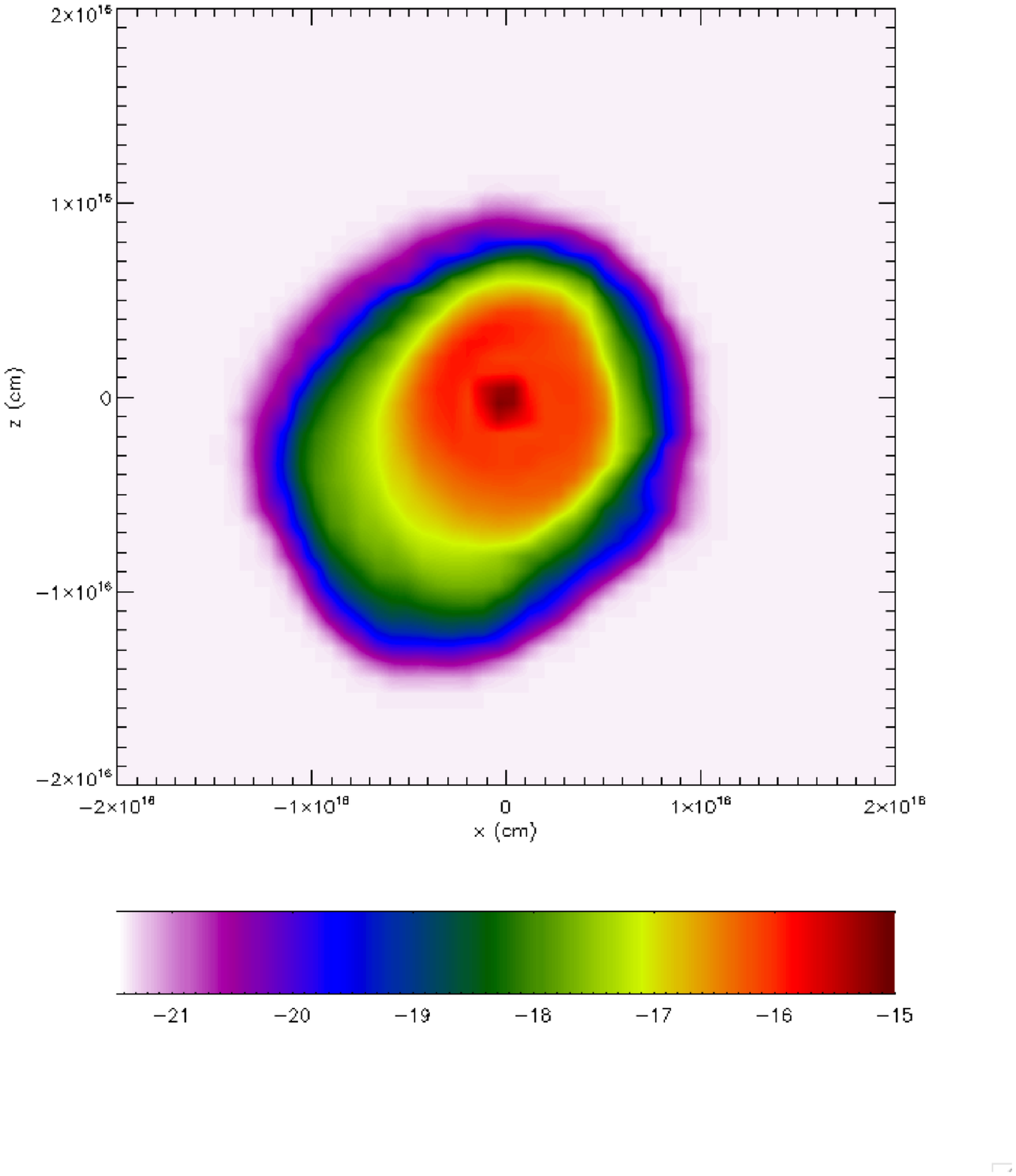}
\vspace{-0.5in}
\caption{Model O log density cross-section in the disk
midplane ($y = -1.77 \times 10^{17}$ cm) at 0.121 Myr. 
At this time, the single sink particle is located at 
$x = -4.8 \times 10^{13}$ cm and $z = -2.3 \times 10^{14}$ cm, 
near the center of the disk.}
\end{figure}
\clearpage

\begin{figure}
\vspace{-1.0in}
\includegraphics[scale=.80,angle=90]{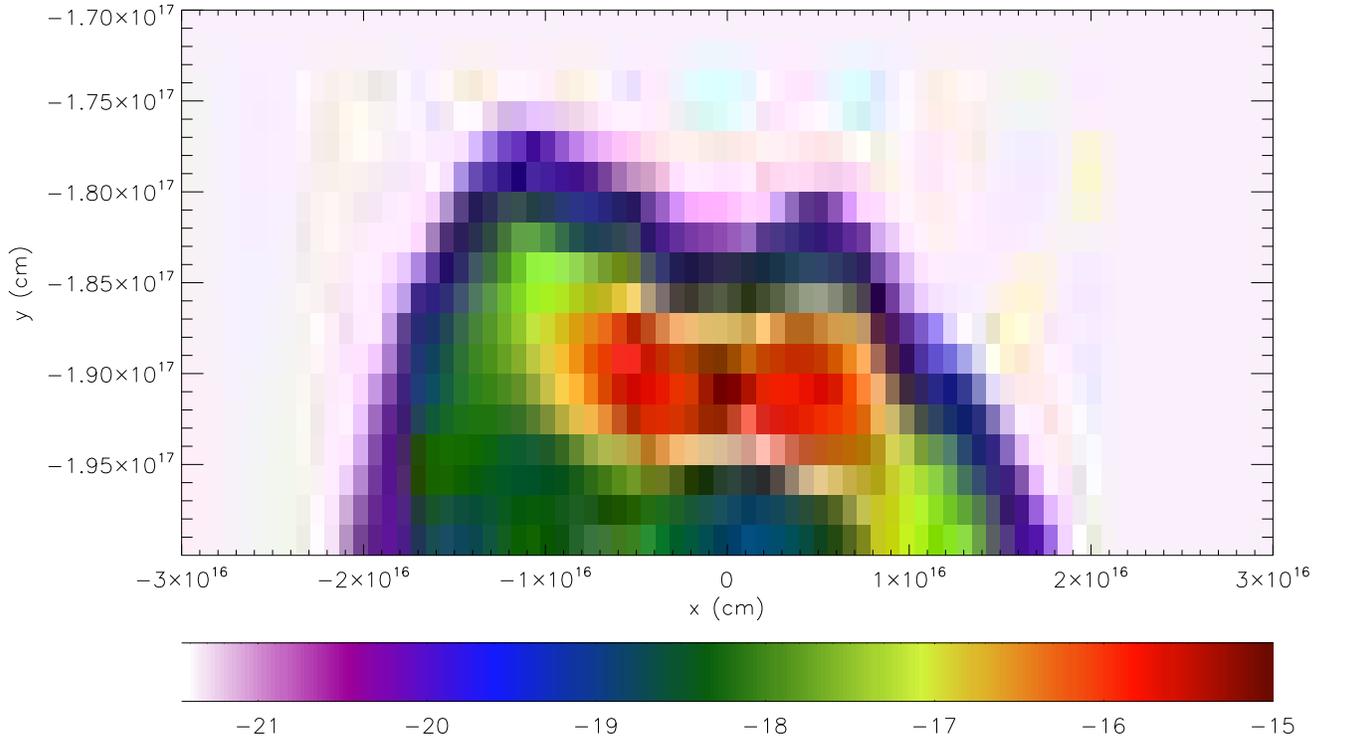}
\vspace{-0.5in}
\caption{Model P log density cross-section ($z$ = 0) at 0.124 Myr. 
At this time, the single sink particle is located at 
$x = -4.3 \times 10^{14}$ cm and $y = -1.91 \times 10^{17}$ cm, 
in the center of the disk midplane.}
\end{figure}
\clearpage

\begin{figure}
\vspace{-1.0in}
\includegraphics[scale=.80,angle=90]{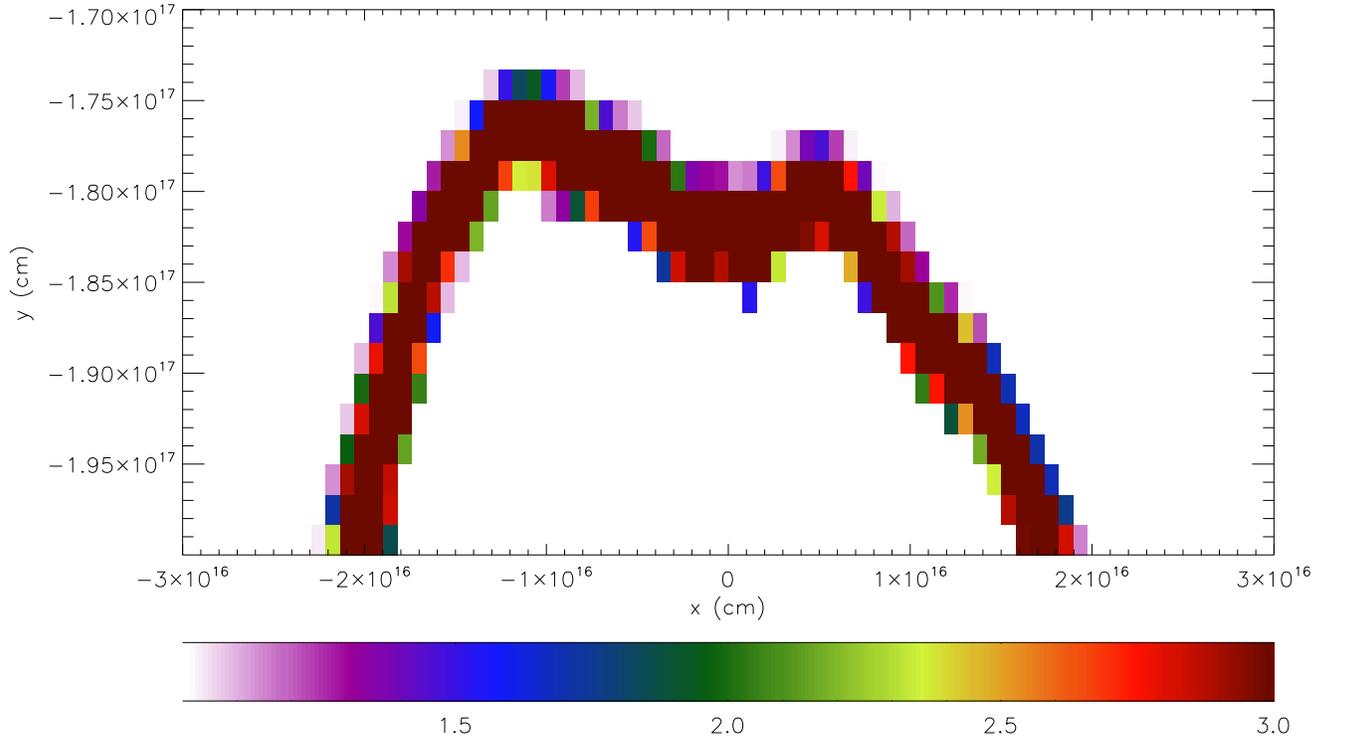}
\vspace{-0.5in}
\caption{Model P log temperature cross-section ($z$ = 0) at 0.124 Myr. 
The high temperature regions are limited to the shock-cloud interface.
The disk remains at 10 K, as the densest gas has been accreted by the
unseen protostar located near $x = -4.3 \times 10^{14}$ cm
and $y = -1.91 \times 10^{17}$ cm.}
\end{figure}
\clearpage

\begin{figure}
\vspace{-1.0in}
\includegraphics[scale=.80,angle=90]{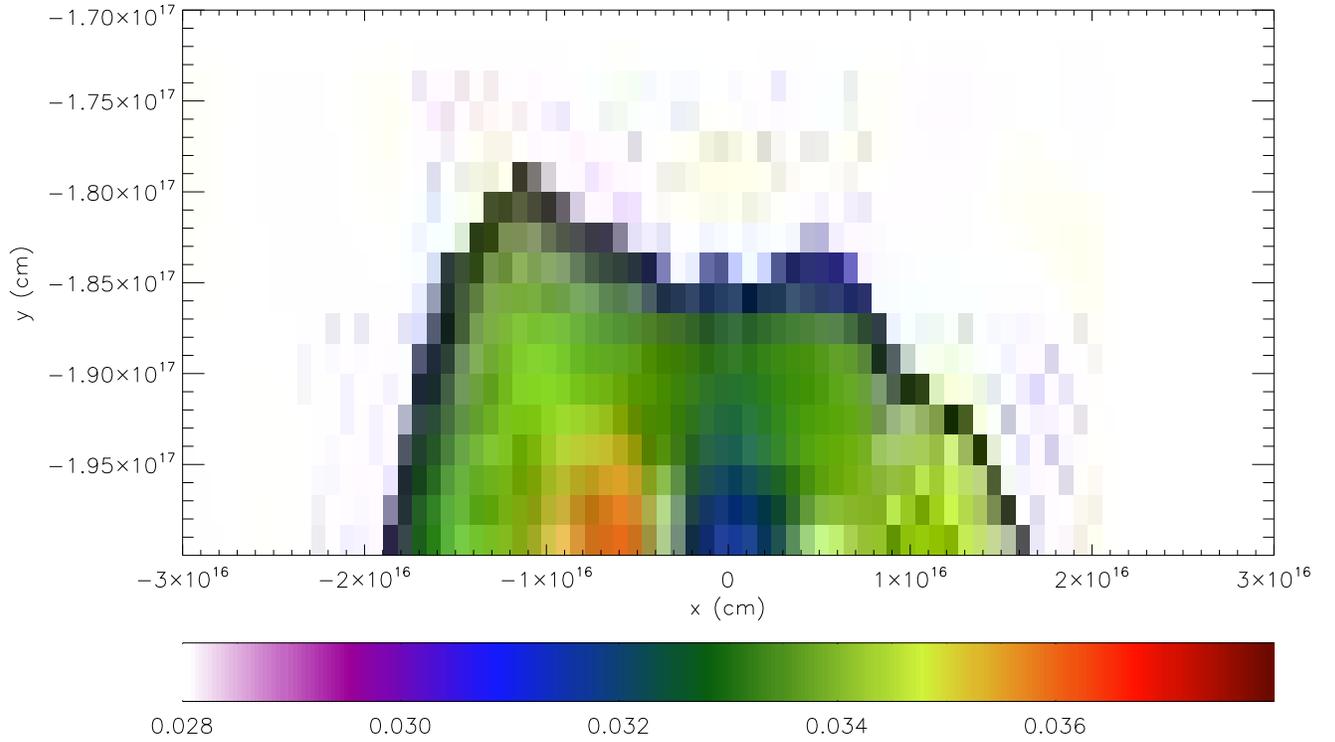}
\vspace{-0.5in}
\caption{Model P color field cross-section ($z$ = 0) at 0.124 Myr. 
The color field represents the space density of SLRIs injected from the
shock front, starting from an initial color field density of one.
At this phase, the SLRIs have been diluted to levels of $\sim 0.034$, but
still pollute the disk and surrounding cloud remnants.}
\end{figure}
\clearpage

\begin{figure}
\vspace{-1.0in}
\includegraphics[scale=.80,angle=0]{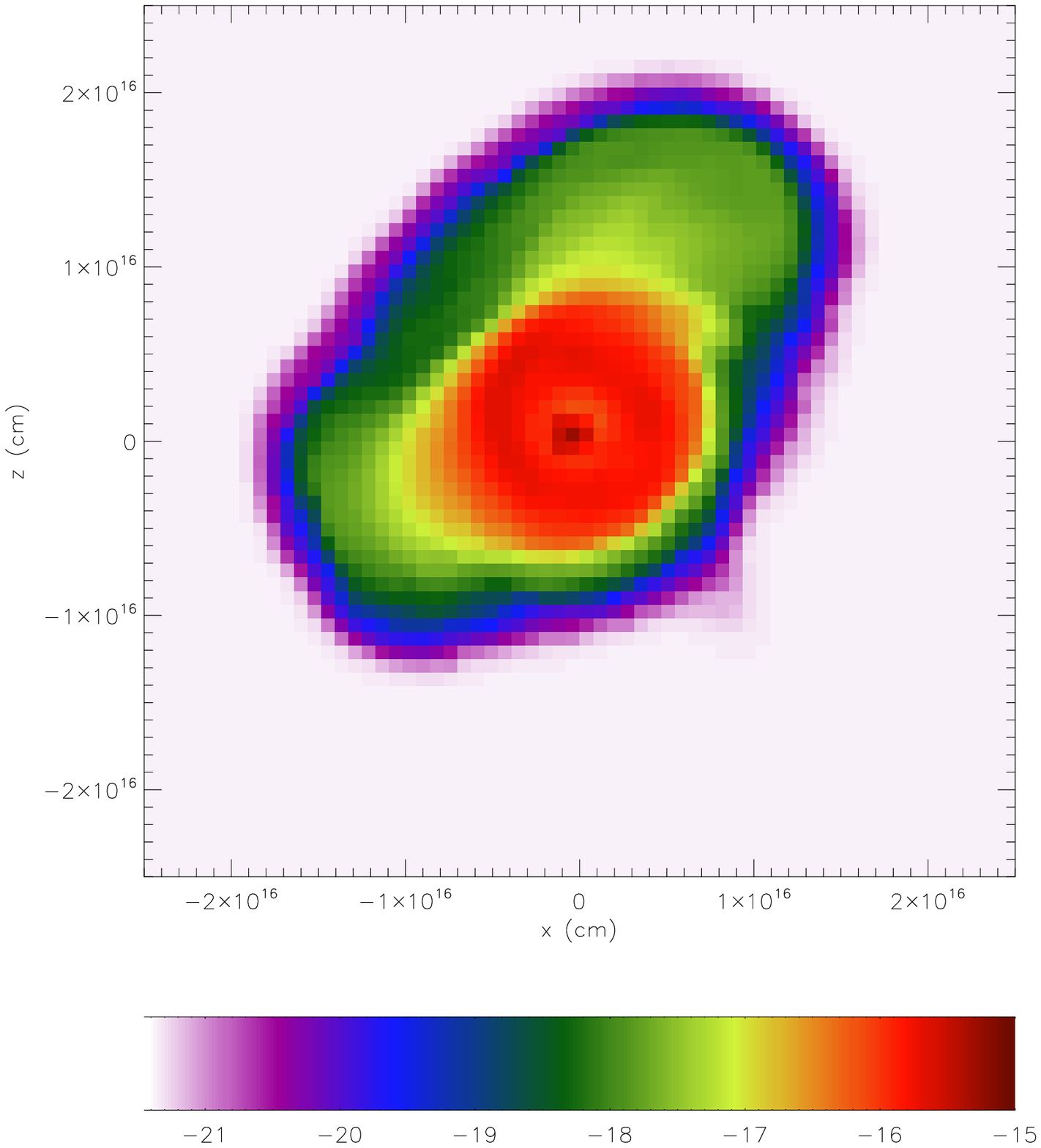}
\vspace{-0.5in}
\caption{Model P log density cross-section in the disk
midplane ($y = -1.91 \times 10^{17}$ cm) at 0.124 Myr. 
At this time, the single sink particle is located at 
$x = -4.3 \times 10^{14}$ cm and $z = 7.1 \times 10^{14}$ cm, 
near the center of the disk.}
\end{figure}
\clearpage

\begin{figure}
\vspace{-1.0in}
\includegraphics[scale=.80,angle=0]{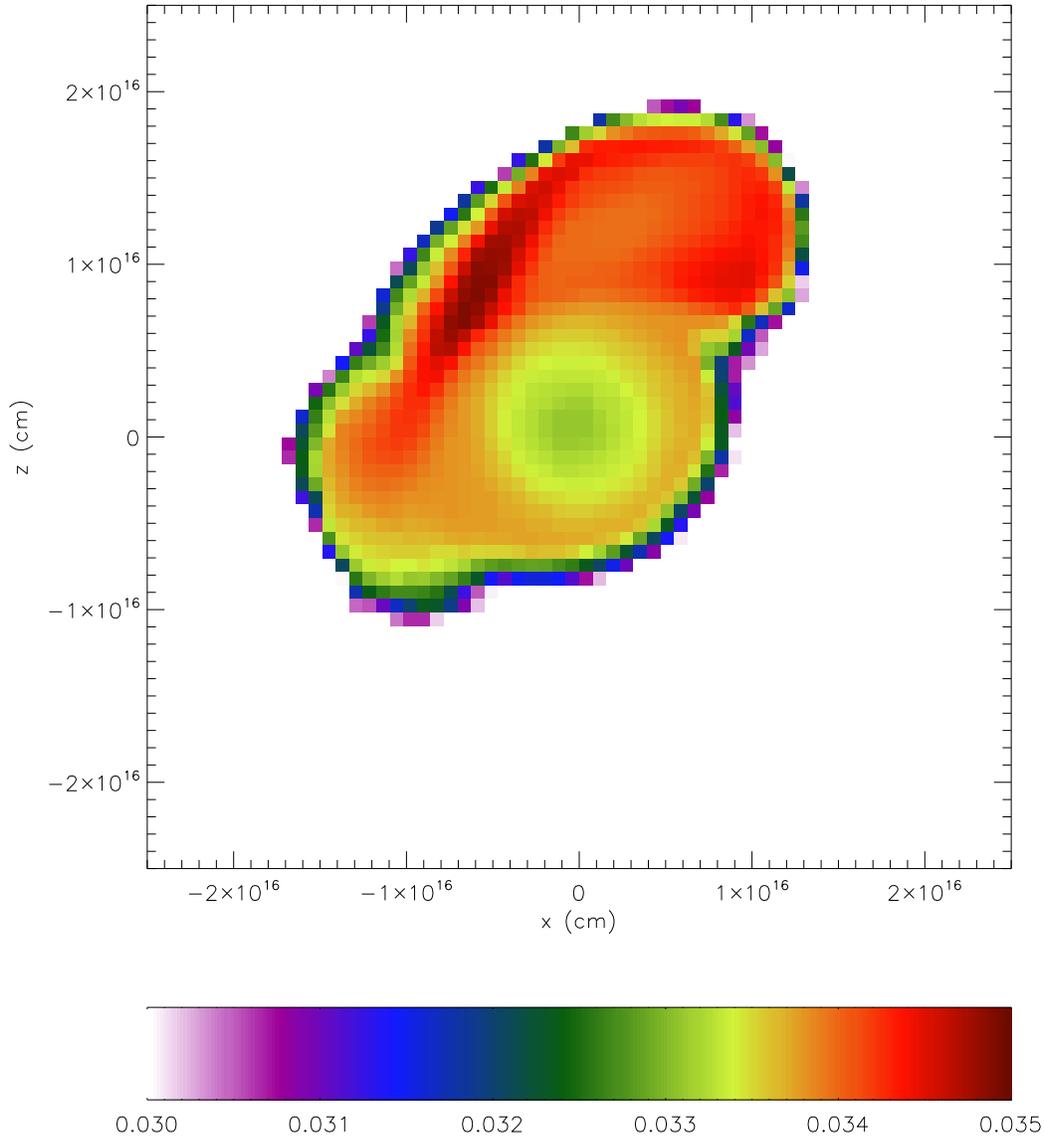}
\vspace{-0.5in}
\caption{Model P color field cross-section in the disk
midplane ($y = -1.91 \times 10^{17}$ cm) at 0.124 Myr. 
Subtle variations in color field density are seen throughout
the disk, a reflection of their injection through a limited
number of Rayleigh-Taylor fingers (e.g., Boss \& Keiser 2014).}
\end{figure}
\clearpage

\begin{figure}
\vspace{-1.0in}
\includegraphics[scale=.80,angle=0]{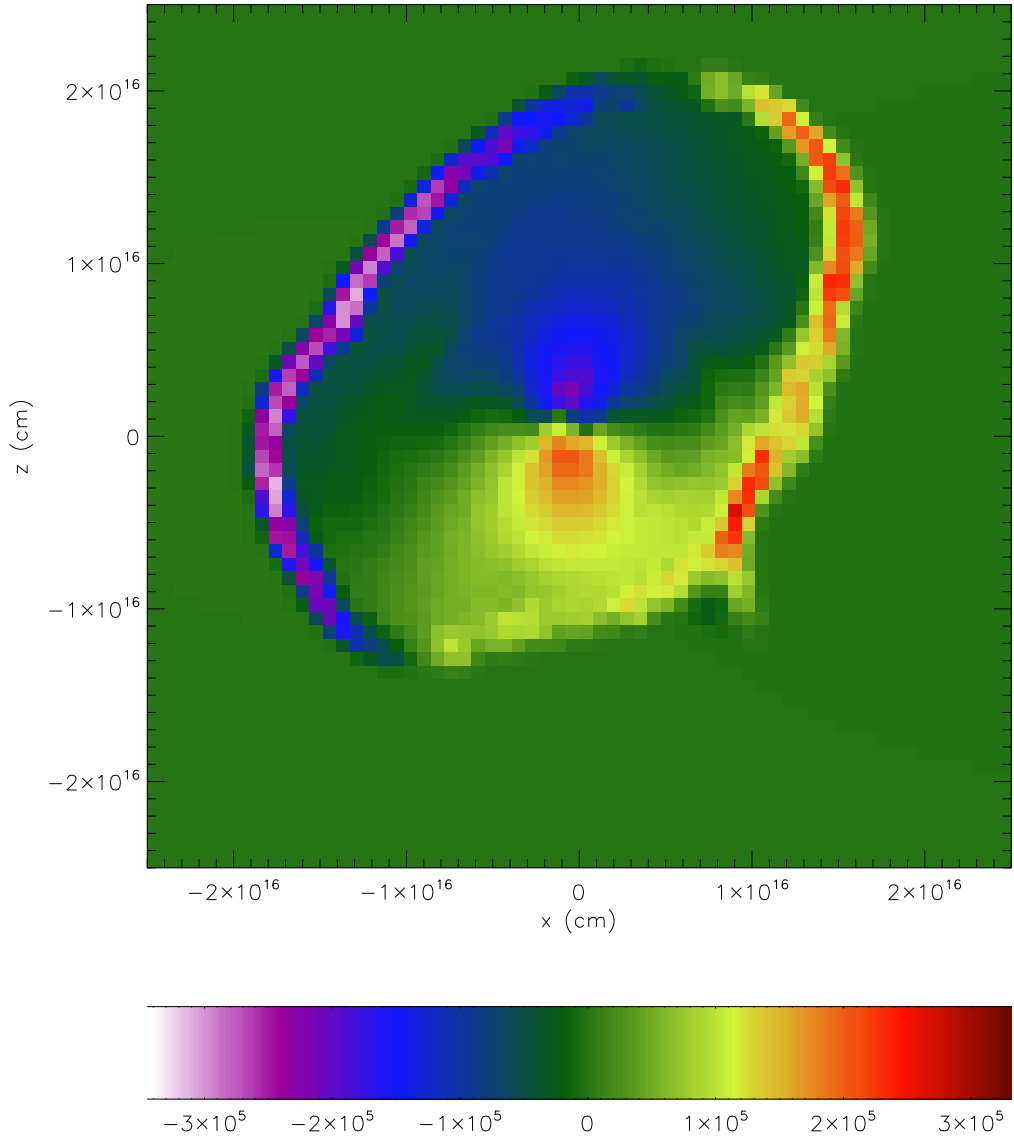}
\vspace{-0.5in}
\caption{Model P $x$ velocity field in the disk
midplane ($y = -1.91 \times 10^{17}$ cm) at 0.124 Myr,
consistent with rotation of the disk about the central sink particle.}
\end{figure}
\clearpage

\end{document}